%% file: main.tex
\pdfoutput=1

\documentclass[11pt]{article}

\input{include/0.package}

\input{include/1.command}

\input{include/2.math}

\usepackage[]{acl}

\usepackage{times}
\usepackage{latexsym}

\usepackage[T1]{fontenc}

\usepackage[utf8]{inputenc}

\usepackage{microtype}

\title{
TOP:A New Target-Audience Oriented Content Paraphrase Task
}

\author{
    Boda Lin$^{1\ddagger}$, Jiaxin Shi$^{2\ddagger}$, Haolong Yan$^1$, Binghao Tang$^1$, Xiaocheng Gong$^1$, Si Li$^{1*}$ \\
    $^1$School of Artificial Intelligence, Beijing University of Posts and Telecommunications\\
    $^2$Huawei Cloud Computing Technologies \\
  \texttt{\{linboda, lisi\}@bupt.edu.cn} 
   }

\begin{document}
\maketitle

\input{section/0.abstract}
\input{section/1.introduction}
\input{section/2.related}
\input{section/3.preliminary}

\input{section/4.method}

\input{section/5.dataset}
\input{section/6.experiment}
\input{section/7.conclusion}
\bibliography{custom}
\clearpage

\appendix

\end{document}

%% file: include/0.package.tex
\usepackage{url}
\usepackage{subfigure}
\usepackage{multirow}
\usepackage{enumitem}
\usepackage{stmaryrd}
\usepackage{array}
\usepackage{threeparttable}
\usepackage{blindtext}
\usepackage{amssymb}

\usepackage[ruled,vlined,linesnumbered,longend]{algorithm2e}
\usepackage{svg}
\usepackage{placeins}

\usepackage{soul}
\usepackage[utf8]{inputenc}
\usepackage{graphicx} 
\usepackage{amsmath}
\usepackage{booktabs}
\usepackage{lipsum}

\usepackage{threeparttable}
\usepackage{makecell}
\usepackage{booktabs}
\usepackage{multirow}
\usepackage{marvosym}
\usepackage{tikz}
\usepackage{amsfonts,amssymb}

%% file: include/1.command.tex
\newcommand{\hide}[1]{}

\newcommand{\task}{Target-Audience Oriented Content Paraphrase~}

%% file: include/2.math.tex
\usepackage{amsmath,amsfonts,bm}

\def\eqref#1{equation~\ref{#1}}

\def\1{\bm{1}}

\DeclareMathAlphabet{\mathsfit}{\encodingdefault}{\sfdefault}{m}{sl}
\SetMathAlphabet{\mathsfit}{bold}{\encodingdefault}{\sfdefault}{bx}{n}



%% file: section/0.abstract.tex
 \begin{abstract}
Recommendation systems usually recommend the existing contents to different users.
However, in comparison to static recommendation methods, a recommendation logic that dynamically adjusts based on user interest preferences may potentially attract a larger user base.
Thus, we consider paraphrasing existing content based on the interests of the users to modify the content to better align with the preferences of users.
In this paper, we propose a new task named Target-Audience Oriented Content Paraphrase aims to generate more customized contents for the target audience.
We introduce the task definition and the corresponding framework for the proposed task and the creation of the corresponding datasets.
We utilize the Large Language Models~(LLMs) and Large Vision Models~(LVMs) to accomplish the base implementation of the TOP framework and provide the referential baseline results for the proposed task.
\end{abstract}

%% file: section/1.introduction.tex
\section{Introduction}

The \task task aims to generate more customized contents for the target audience of the recommendation systems.

Current Recommendation Systems recommend the existing contents to different users depending on the user preferences.
However, some content may also be of interest to the target audience after paraphrasing corresponding to the user preference.
Considering the outstanding generation ability of the current Large Language Models~(LLMs) and Large Vision Models~(LVMs), we have a question: Can we achieve content paraphrasing for the target audience via these large models?

LLMs such as ChatGPT\footnote{https://openai.com/blog/chatgpt}, LLaMa~\cite{touvron2023llama} and LVMs such as Diffusion~\cite{ho2020ddpm}, LLaVA~\cite{liu2023llava} achieve the astonishing results in Natural Language Processing~(NLP) and Computer Vision~(CV) fields.
These models not only achieve state-of-the-art~(SOTA) performance in various NLP and CV tasks but also demonstrate sufficient generalization and ease of use in user interaction scenarios~\cite{touvron2023llama, liu2023llava, zhu2023minigpt, zeng2022glm}.

Thus, we propose a new task named \textbf{T}arget-Audience \textbf{O}riented Content \textbf{P}araphrase (\textbf{TOP}) and explore whether the basic implementation of the proposed TOP task can be achieved through the generation capabilities of LLMs and LVMs.

\begin{figure}[t]
\centering
\includegraphics[width=1.0\linewidth]{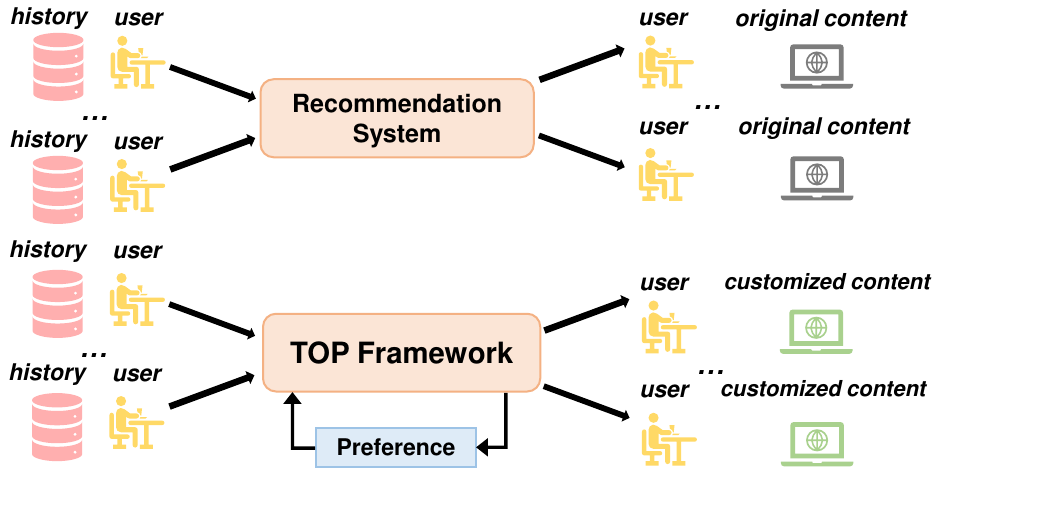}
\caption{
Compared to the approach of directly recommending content to users, the proposed TOP task places a greater emphasis on appropriately paraphrasing existing content based on the preferences of users.
}
\label{fig:intro:top}
\end{figure}
The TOP task aims to generate customized outputs for the current content inputs depending on specific user history.
We design a TOP framework to accomplish the TOP task.
The TOP framework includes the data pre-processing, preference extractor, preference encoder, content encoder, and the content generator.
We also provide corresponding metric designs to measure the quality of generation from three perspectives: the consistency between the generated results and the original content, the consistency between the generated results and preferences, and the quality of the generated results themselves.

In the proposed TOP framework, we supply the model with historical data of the user and require the model to deduce user preferences from these data. 
Subsequently,  the model will paraphrase its current output based on the summarized user preferences. 
This approach aims to tailor the responses of the model more closely to individual user preferences, leveraging historical interaction data to inform and adjust its output.

To facilitate the research about the preference-oriented paraphrase, we collect the data from the Internet and create a Customized-TOP dataset, including $46,372$ texts and $39,685$ images, covering Chinese and English.

We also provide the base implementation of the TOP framework for the text modality and image modality.
To evaluate the performance of the TOP framework and other possible models, we propose Content Preservation Index~(CPI), Preference Preservation Index~(PPI), and Natural Realism Index~(NRI) metrics to measure the model performance from different dimensions.

%% file: section/2.related.tex
\section{Related Work}
\subsection{LLMs}
Large Language Models~(LLMs) have become pivotal in the field of NLP. 
Starting from foundational models like BERT~\cite{kenton2019bert} and GPT-2~\cite{gpt2}, and advancing to more recent iterations such as GPT-3~\cite{gpt3}, InstructGPT~\cite{instructgpt}, as well as other open-source, large-scale language models including LLaMA~\cite{touvron2023llama} and LLaMA2~\cite{llama2}, significant progress has been made in NLP, particularly in natural language understanding and generation.
BLIP~\cite{blip} has further advanced multimodal capabilities by pre-training a mixture of encoder-decoder models to enhance vision-language tasks.

\subsection{Diffusion Models}
In the field of Computer Vision, diffusion models~\cite{sohl2015deep} have achieved great improvements in various tasks, especially in image generation.
Diffusion models first perturb the data to a standard Gaussian noise by gradually adding noise into the data, which is formulated as a Markov chain.
The data can be denoised through the reverse procedure.

The latent diffusion model~\cite{rombach2022high} proposes generating images by iteratively denoising data in a latent representation space, significantly reducing the computational burden of the diffusion model. Following that, several efficient training methods emerged to fine-tune custom diffusion models, including Textual inversion~\cite{gal2022image}, Dreambooth~\cite{ruiz2022dreambooth}, and IP-adapter~\cite{ye2023ip-adapter}. These adapters leverage a series of image inferences and trigger words to generate specific objects or styles.

\subsection{Recommendation System}
Content-based recommendation system~\cite{pazzani2007content} represents a mainstream method of recommendation. 
It aims to suggest products to users based on the attributes of the content and the historical preferences of users.
As content-based recommendation systems do not require the consideration of the item-user interaction matrix and do not necessitate data from other users to generate recommendations, they are widely applied across various fields.

\citet{shu2018contentcnn} propose using Convolutional Neural Network~(CNN) and Bregman iteration method to achieve content-based text recommendation without pre-labeling.
MRS~\cite{yao2015mrs} using a content-based recommendation system to recommend movies and applied in the streaming platforms.
IALS~\cite{dhawan2024novel} improves the alternating least square algorithm and combines the stochastic gradient descent algorithm to achieve content-based recommendations for social media.

%% file: section/3.preliminary.tex
\section{Preliminaries}

\begin{figure*}[t]
\centering
    \includegraphics[width=1.0\linewidth]{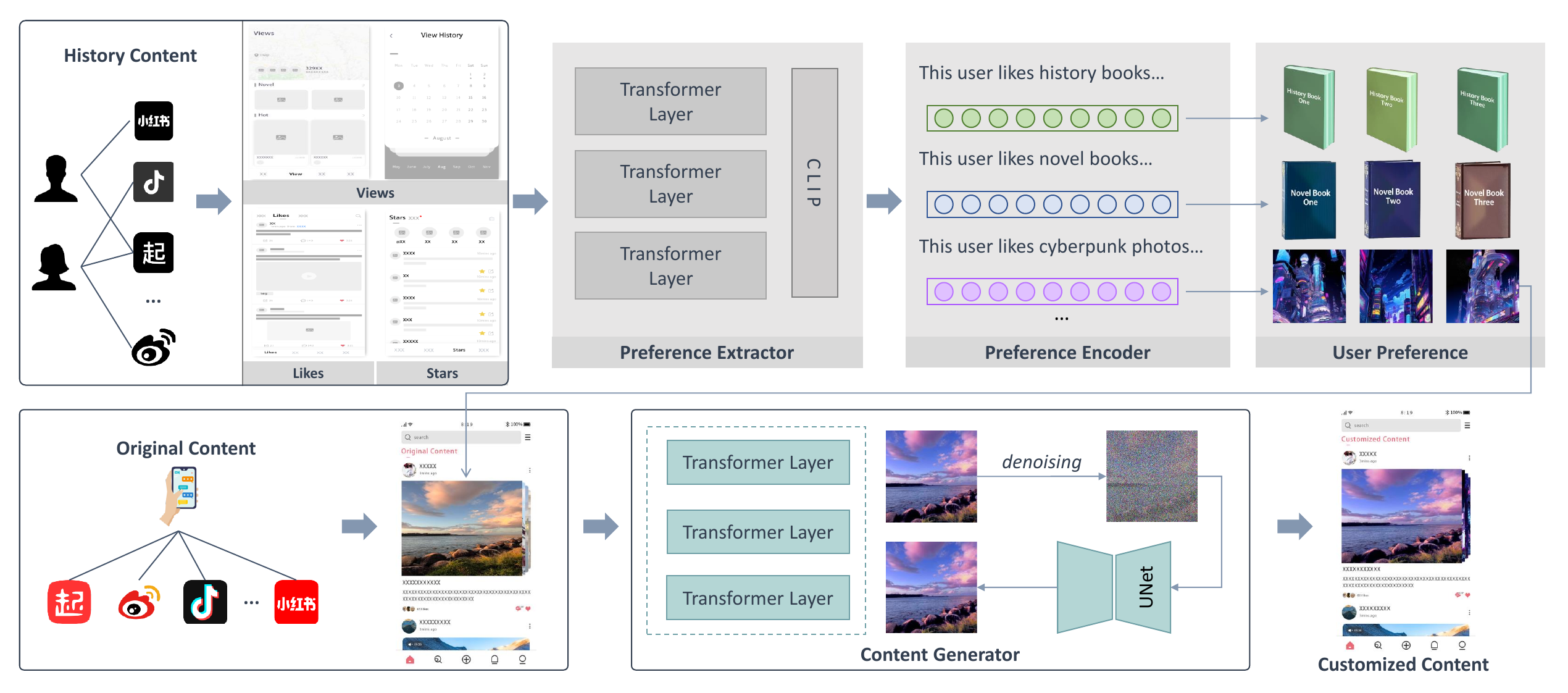}
\caption{
The total framework of the proposed TOP task.
The user preference is extracted by the preference extractor and encoded by the preference encoder.
And the final customized output is generated by the content generator.
}
\label{fig:task}
\end{figure*}

\subsection{Task Definition}

We formally introduce the proposed Target-Audience Oriented Content Paraphrase task and the corresponding notations.
This paper uses bold lower case letters, blackboard letters, and bold upper case letters to denote sequences, sets, and functions, respectively.
Elements in the sequence and the sets are enclosed in parentheses and braces, respectively.

The TOP task aims to summarize the preference of the specific user from user history and paraphrase the current input content depending on the summarized preference, which makes the paraphrased output more in line with the user preference.
The formal definition of the TOP task is
\begin{equation*}
    \min ||O-P|| + ||O-C||
\end{equation*}

The TOP task includes the following elements:${H, C, P, O}$, where $H = \left[h_1, h_2, \dots, h_n\right]$ is the historical data list of the current user, $C$ is the current input content.
The history data and the input data may come from multiple modalities such as text, image, video and audio, etc.
The extracted preference $P$ also contains various forms, which may be natural language, image collection, implicit vector representation, etc. 
The paraphrased result $O$ is consistent in representation form with the content data $C$.

\subsection{Unified Framework}

As shown in Fig~\ref{fig:task}, we also provide a TOP framework corresponding to the proposed TOP task.
The TOP framework includes the following modules: the preference extractor $P_{ext}$, which responds to extracting user preference from the user historical data. The preference encoder $P_{enc}$, which responds to encoding the extracted preference. The content encoder $C_{enc}$, which responds to encoding the current input content. The content generator $C_{gen}$, which responds to generating the paraphrased output.

A comprehensive TOP framework encompasses the following stages: 1) Data collection and pre-processing. 
2) Training of the preference model and extraction of user preferences from historical data. 
3) Training of the paraphrase model and performing content paraphrasing based on user preferences and current input.
4) Selecting appropriate metrics for performance evaluation based on the modality of the data.

In the preference extracting stage, the preference extractor $P_{ext}$ extracts user preference based on the user specific historical data, and the preference encoder $P_{enc}$ encodes the preference features.

\begin{equation*}
    P = \text{$P_{enc}$}(\text{$P_{ext}$}(H))
\end{equation*}
In the content paraphrase stage, the content generator $G$ generates the paraphrased output $O$ based on the extracted preference $P$ and the content input $C$.
\begin{equation*}
    O = \text{$C_{gen}$}(\text{$C_{enc}$}(C)\mid P)
\end{equation*}

\subsection{Metric Design}
\label{sec:metric_design}
To evaluate the performance of the model in the proposed TOP task, we propose the TOP task should include three categories of metrics, namely
\textbf{C}ontent \textbf{P}reservation \textbf{I}ndex~(\textbf{CPI}), \textbf{P}reference \textbf{P}reservation \textbf{I}ndex~(\textbf{PPI}), and \textbf{N}atural \textbf{R}ealism \textbf{I}ndex~(\textbf{NRI}).
The CPI metrics are designed to assess the consistency between the outputs of the model and the original inputs.
The formal definition is as follows:
$$
\text{CPI} = \mathrm{F}_c(\mathrm{O}, \mathbf{C})
$$
where the $\mathrm{F}_c$ is a function used to measure the consistency between the output $O$ and the input content $C$.
These metrics are used to verify whether the content diverges excessively from the semantic context of the original input after paraphrasing. 
The PPI metrics are intended to measure the alignment of the output of the model with user preferences. 
This aims to validate whether the paraphrased content aligns with preferences as indicated by the user historical data.
The formal definition is as follows:
$$
\text{PPI} = \mathrm{F}_c \left(\mathrm{O}, \mathrm{P}\right)
$$
The NRI metrics are employed to evaluate the intrinsic quality of the outputs, such as the fluency and fidelity of the generated results.
The formal definition is as follows:
$$
\text{NRI} = \mathrm{F}_q\left(\mathrm{O}\right)
$$
where the $\mathrm{F}_q$ is a function to measure the natural quality of the output.

%% file: section/4.method.tex
\section{Implementation}

\begin{figure*}[t]
\centering
    \includegraphics[width=1.0\linewidth]{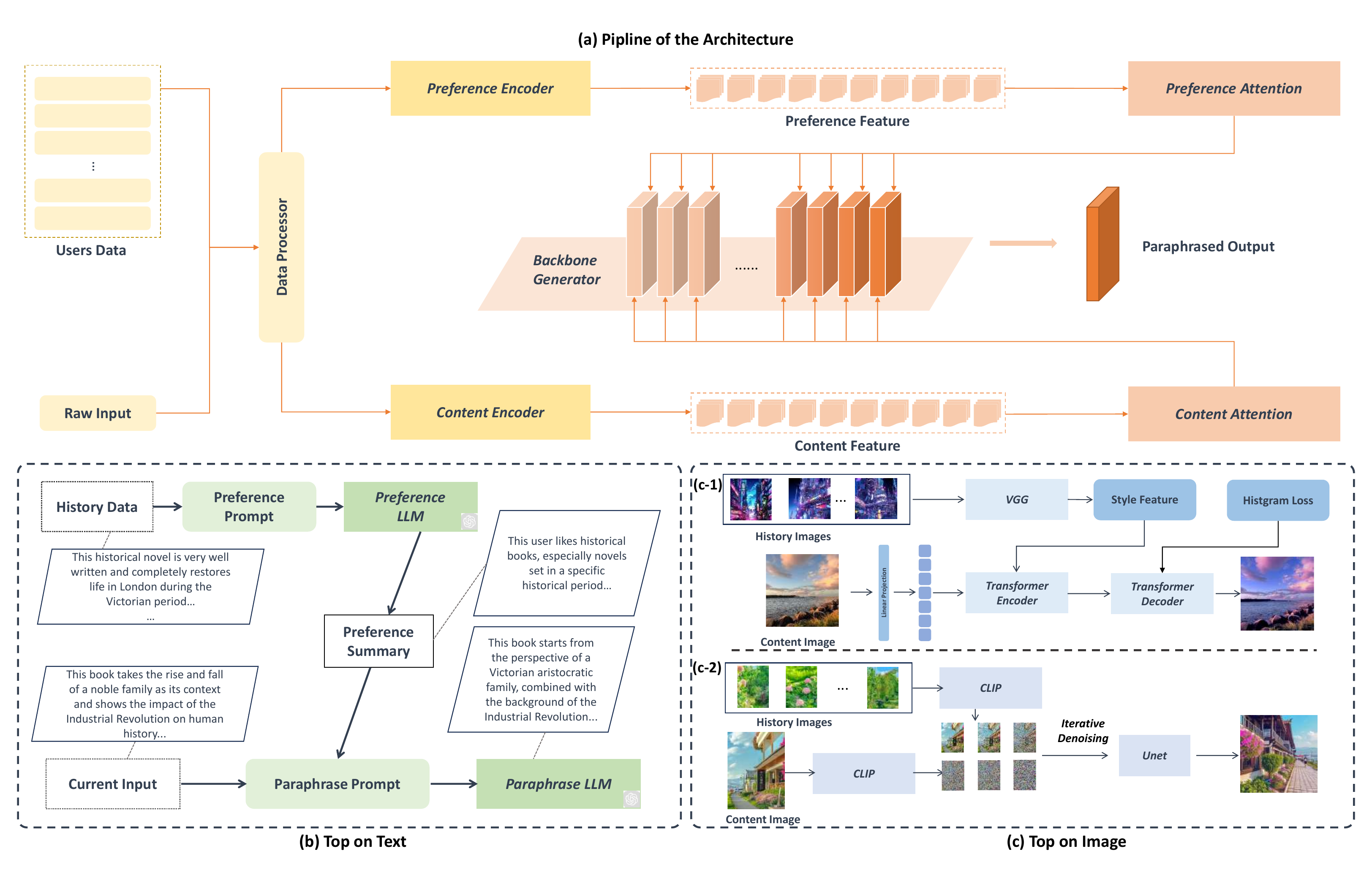}
\caption{
The figure of our base implementation of the TOP framework.
Panel (a) illustrates the overall framework, which aligns with the TOP framework described in Section 3. 
Panel (b) delineates the specific implementation of TOP for text. 
Historical data is concatenated with a preference prompt and then fed into a preference LLM to extract the preferences. 
The content text is concatenated with a paraphrase prompt and in conjunction with the extracted preferences is processed through a paraphrase LLM to generate the outcome.
Panel (c) demonstrates the specific implementation of TOP for images.
(c-1) depicts the implementation of TOP applied to the task of image style transfer and (c-2) depicts the implementation of TOP applied to the task of iamge concept transfer.
}
\label{fig:model}
\end{figure*}

In this section, we provide a detailed exposition of the base implementation for the proposed TOP task.
In this paper, our primary focus is on the text and image modalities.
Specifically, for the image modality, we consider implementing the TOP task from two distinct perspectives: image style and image concept.

\subsection{TOP on Text}

For the implementation of TOP on text, we employ the LLM as our preference extractor $P_{ext}$ and content generator $C_{gen}$, which is composed of multiple layers of Transformer decoders.
In the preference extracting stage of the TOP task, we utilize LLM to summarize the preference of the target user.
The historical text $H = {\langle h_1, h_2, \dots, h_n \rangle}$, where $h_i$ is the $i$-th text sequence data.
We concatenate a pre-designed instruction template $I^{p}$ to the historical sequence and obtain the final historical sequence $H' = {\langle I^{p} \oplus H\rangle}$.
The preference is extracted and summarized by the $P_{ext}$ in the form of natural language.
\begin{equation*}
    P = P_{ext}(H')
\end{equation*}
The preference feature $P$ is concatenated with the current content input $x$ and the paraphrase prompt $I^{c}$: $x'={\langle I^{c} \oplus P \oplus x \rangle}$.
The final paraphrased content output is generated by the $C_{gen}$.
\begin{equation*}
    O = C_{gen}(x')
\end{equation*}
The loss of the training stage is the commonly used next token prediction loss of LLMs:
$$\mathcal{L} = -\sum_{i=1}^{n} \log p_{\theta}(y_i | x'_1, x'_2, ..., x'_{i-1})) $$
where $\theta$ represents the parameters of LLM and $x'_i$ is the $i$-th token of final input sequence $x'$.

\subsection{TOP on Image}
As shown in Figure~\ref{fig:model}, we decompose the TOP task on images into two sub-tasks: style transfer and concept transfer.

\textbf{TOP for Image style.}
We use the gram matrix to characterize the style features of an image and employ the VGG  model~\cite{simonyan2014vgg} as $P_{ext}$ and $P_{enc}$ to extract the gram matrix following previous research~\cite{gatys2015texture, gatys2015neural, deng2022stytr2}.
We also employ the vision Transformer as content encoder $C_{enc}$ and content generator $C_{gen}$.

In the each layer $l$ of the VGG, the style loss $L^{l}_{sty}$ is calculated as:
\begin{equation*}
    L_{style}^l(G, A) = \frac{1}{4N_l^2M_l^2} \sum_{i,j} (G^l_{ij} - A^l_{ij})^2
\end{equation*}
where $G^l$ denotes the gram matrix for the generated image and $A^l$ for the style image, $N_l$ is the number of feature maps and $M_l$ is the dimensions of each feature map as layer $l$.

The overall style loss $L_{sty}$ is the weighted sum of the style losses across multiple layers:
\begin{equation*}
    L_{style}(G, A) = \sum_l w_l \cdot L_{style}^l(G, A)
\end{equation*}
where $w_l$ denotes the weight assigned to the style loss as layer $l$.

The content loss $L_{content}$ is formulated as:
\begin{equation*}
    L_{content}(F, P) = \frac{1}{2} \sum_{i, j} (F_{ij} - P_{ij})^2
\end{equation*}
where $F_{ij}$ and $P_{ij}$ correspond to the elements in the feature maps of the generated and content images, respectively.

We also use the Radial Basis Funciton~(RBF) kernel function to measure the difference between the histogram feature of input image $hist^{cur}$ and history image $hist^{his}$.
\begin{equation*}
    L_{hist} = \exp{(-\gamma{\Vert {hist^{cur} - hist^{his}} \Vert_2})}
\end{equation*}

Finally, the total loss function is a weighted combination of the content and style losses:
\begin{equation*}
    L_{total} = \alpha \cdot L_{content}(F, P) + \beta \cdot L_{style}(G, A)
\end{equation*}
Here, $\alpha$ and $\beta$ are parameters that balance the importance of content and style, respectively.

\textbf{TOP for Concept Transfer.}
Different from style transfer, concept transfer focuses more on transferring some major concepts in historical images to content images.
We achieve this task based on the Diffusion method.
We employ the CLIP~\cite{clip} as preference extractor $P_{ext}$, preference encoder $P_{enc}$, and content encoder $C_{enc}$.
The UNet is employed as content generator $C_{gen}$.

The diffusion model includes two main processes: the forward process~(diffusion process) and the reverse process~(denoising process).
During the forward process, the diffusion model adds Gaussian noise to the original data via the Markov chain of T steps.
After enough steps, the data becomes nearly indistinguishable from pure noise.
In the reverse process, the diffusion model learns to generate the samples from the noise generated by the forward process.
Some inputs such as text can be injected into the diffusion model as a condition to guide the reverse process.

Different from the traditional text-condition diffusion models, we use history images of the user as the condition to guide the reverse process.
The preference feature is extracted by the image encoder and sent into the Unet.

The training objective of the diffusion model, denoted as $\epsilon_{\theta}$, which predicts noise, is defined as a simplified variant of the variational bound:

$L_{\text{simple}} = \mathbb{E}_{x_0, \mathbf{\epsilon} \sim \mathcal{N}(0,\mathbf{I}), c, t} \left\| \mathbf{\epsilon} - \mathbf{\epsilon}_{\theta} (x_t, c, t) \right\|^2
$
where \(x_0\) represents the real data with an additional condition \(c\), \(t \in [0, T]\) denotes the time step of diffusion process, \(x_t = \alpha_t x_0 + \sigma_t \epsilon\) is the noisy data at \(t\) step, and \(\alpha_t\), \(\sigma_t\) are predefined functions of \(t\) that determine the diffusion process.

%% file: section/5.dataset.tex
\section{Datasets}
The proposed Customized-TOP dataset includes the text modality and image modality.
We collect these data from the Internet. 
All the data utilized in this paper have been subjected to anonymization and encryption processes to ensure privacy and security compliance.
We employ our proprietary numbering system to reassign user identifiers (userids) to all users.
Then we utilize the Advanced Encryption Standard~(AES) Algorithm~\cite{daemen1999aes} for userids to achieve anonymization.

\begin{figure}[t]
\centering
\includegraphics[width=1.0\linewidth]{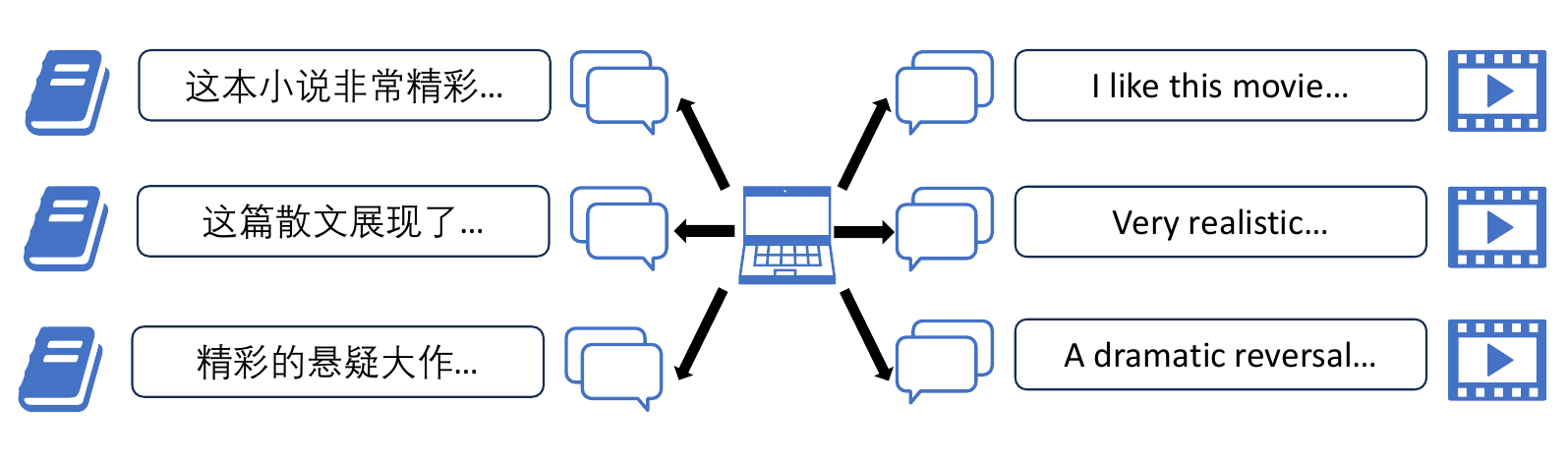}
\caption{
Display of the original source of text data.
}
\label{fig:data:website:text}
\end{figure}

\subsection{Text Datasets}
We implement the TOP on text in the scenario of the books and movies introduction paraphrasing.
As shown in Figure~\ref{fig:data:website:text}, we independently collect bilingual corpora consisting of English and Chinese texts, specifically from websites that publish movie and book comments.
These comments exhibit user preferences regarding the genres of books and movies to a certain extent.

All comment data is subjected to anonymization and preprocessed into the following format:
$\langle \mathtt{userid}, \mathtt{piecename}, \mathtt{comment}\rangle$, where the $\mathtt{userid}$ is encrypted id and the $\mathtt{piecename}$ refers to the name of the book or movie currently being discussed in the comment.
After extracting the user preference, the preference attribute is appended into data item:
$\langle \mathtt{userid}, \mathtt{piecename}, \mathtt{comment}, \mathtt{preference} \rangle$.

For the TOP task in the second phase, the official introduction of books and movies is incorporated into the dataset. 
Following pre-processing, the data is formatted as follows:
$\langle \mathtt{piecename}, \mathtt{intro}, \mathtt{preference}, \mathtt{output} \rangle$, where the $\mathtt{output}$ refers to the rewriting outputs of introduction depend on the current user preference.
\input{tables/0.data.text}

The statistics details of text datasets are shown in Table~\ref{tab:data_text}, the text data used for paraphrase consists of user preferences cross-matched with different piece introductions.

\subsection{Image Datasets}
\begin{figure*}[t]
\centering
\includegraphics[width=1.0\linewidth]{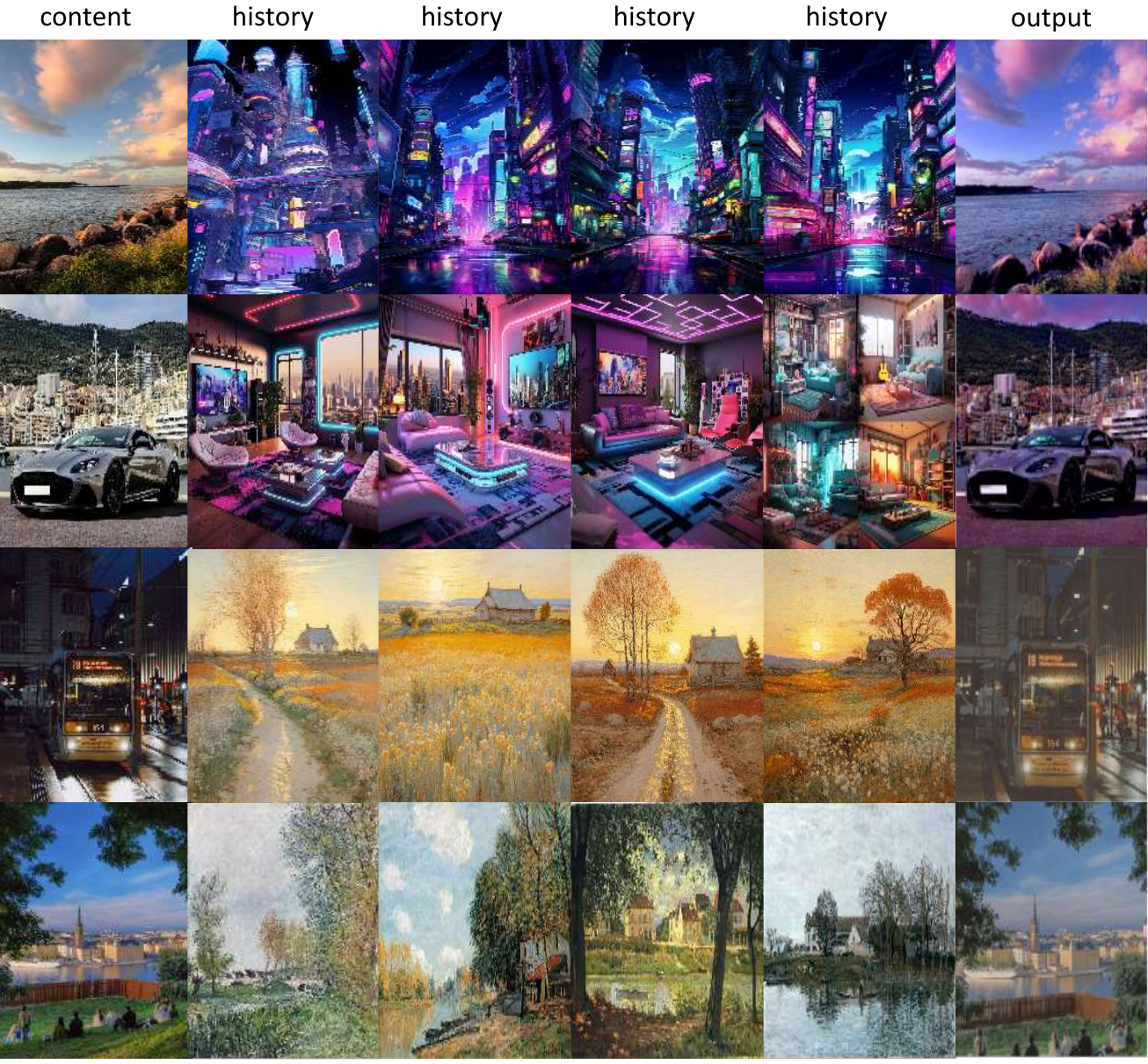}
\caption{
Teasers of TOP on Image for Style Transfer.
It can be observed that due to the histogram loss and the reduction of the weight of the style loss, our generation results do not produce texture distortion which is common in photo style transfer.
}
\label{fig:teaser:result:result}
\end{figure*}
\begin{figure*}[h]
\centering
\includegraphics[width=1.0\linewidth]{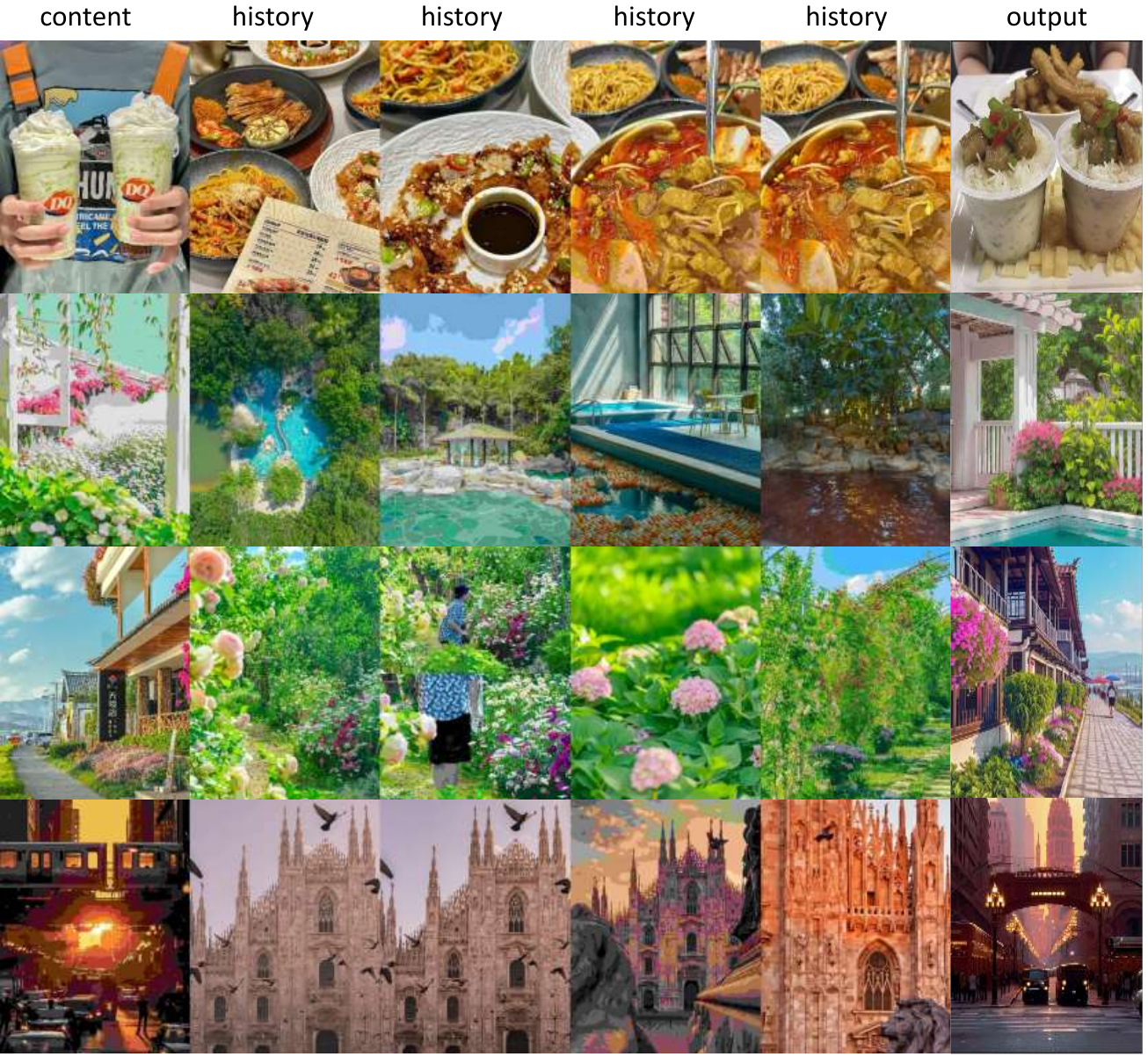}
\caption{
Teasers of TOP on Image for Concept Transfer.
}
\label{fig:teaser:result:concept}
\end{figure*}

For the image data, we conducted a manual screening process to meticulously eliminate images of suboptimal quality.
Then we employ the LAION aesthetic predictor\footnote{https://github.com/LAION-AI/aesthetic-predictor} for quality filtering.
We select images that have a LAION aesthetic score exceeding $5.0$.

\input{tables/1.data.image}
\input{tables/resutls.text}
The statistics details of the proposed image datasets are shown in Table~\ref{tab:data_image}, the image data for paraphrase consists of user preferences cross-matched with different content images.

%% file: tables/0.data.text.tex
\begin{table}[t]
\centering
	\scalebox{0.8}{
	\begin{tabular}{cccccc}
    \toprule
        \multicolumn{2}{c}{\textbf{Dataset}} & \textbf{Users} & \textbf{Pieces} & \textbf{Reviews} & \textbf{Data} \\
        \midrule
         \multirow{3}{*}{English} & train & $2,226$ & $528$ & $27,202$ & $23,885$ \\
         & dev & $330$ & $76$ & $3,849$ & $2,844$ \\
         & test & $639$ & $144$ & $7,440$ & $6,012$ \\
         \midrule
         \multirow{3}{*}{Chinese} & train & $1,638$ & $862$ & $6,531$ & $42,010$ \\
         & dev & $227$ & $98$ & $417$ & $1,207$ \\
         & test & $406$ & $203$ & $933$ & $1,907$ \\
    \bottomrule
	\end{tabular}
  }
	\caption{
The statistics details of the proposed text datasets.
	}
	\label{tab:data_text}
\end{table}

%% file: tables/1.data.image.tex
\begin{table}[t]
\centering
	\scalebox{1.0}{
	\begin{tabular}{ccccc}
    \toprule
        \multicolumn{2}{c}{\textbf{Dataset}} & \textbf{Users} & \textbf{Images} & \textbf{Data} \\
        \midrule
         \multirow{3}{*}{Style} & train & $244$ & $4,140$ & $42,010$ \\
         & dev & $70$ & $264$ & $166$ \\
         & test & $104$ & $568$ & $240$ \\
         \midrule
         \multirow{3}{*}{Concept} & train & $460$ & $31,429$ & $5,112$ \\
         & dev & $40$ & $1,238$ & $300$ \\
         & test & $82$ & $2,046$ & $601$ \\
    \bottomrule
	\end{tabular}

  }
	\caption{
The statistics details of the proposed image datasets.
	}
	\label{tab:data_image}
\end{table}

%% file: tables/resutls.text.tex
\begin{table}[t]
\centering
	\scalebox{0.64}{
	\begin{tabular}{cccccc}
    \toprule
        \multirow{2}{*}{\textbf{Model}} & \multirow{2}{*}{\textbf{Language}} & \multicolumn{2}{c}{\textbf{CPI}} & \multicolumn{1}{c}{\textbf{PPI}}  & \multicolumn{1}{c}{\textbf{NRI}}  \\
         & & Rouge-L & Bert-Score & Bert-Score & PPL \\
         \midrule
         GPT-4 & English & $45.77$ & $78.32$ & $58.29$ & $5.03$\\
         TOP & English & $42.89$ & $76.31$ & $58.60$ & $5.41$ \\
         GPT-4 & Chinese & $43.33$ & $76.10$ & $58.74$ & $5.15$\\
         TOP & Chinese & $42.89$ & $75.42$ & $58.03$ & $6.59$ \\
         \midrule
          \multirow{2}{*}{\textbf{Model}} & \multirow{2}{*}{\textbf{Task}} & \multicolumn{2}{c}{\textbf{CPI}} & \multicolumn{1}{c}{\textbf{PPI}}  & \multicolumn{1}{c}{\textbf{NRI}}  \\
         & & \multicolumn{2}{c}{$\mathcal{L}_s$} & $\mathcal{L}_s$ & AES-Score \\
         \midrule
         TOP & Style Trasnfer &  \multicolumn{2}{c}{$0.65$} & $0.47$ & $6.35$ \\
         \midrule
         \multirow{2}{*}{\textbf{Model}} & \multirow{2}{*}{\textbf{Task}} & \multicolumn{2}{c}{\textbf{CPI}} & \multicolumn{1}{c}{\textbf{PPI}}  & \multicolumn{1}{c}{\textbf{NRI}}  \\
         & & SSIM & CLIP-I & CLIP-I & AES-Score \\
         \midrule
         TOP & Concept Transfer & $0.32$ & $0.79$ & $0.66$ & $6.49$ \\

    \bottomrule
	\end{tabular}
  }
	\caption{
The results of the our base implementation on Customized-TOP dataset.
GPT-4 represents the results when using GPT-4 as the preference LLM and paraphrase LLM.
	}
	\label{tab:result:text}
\end{table}

%% file: section/6.experiment.tex
\section{Experiments}
\subsection{Automatic Metrics}

As described in Section~\ref{sec:metric_design}, we design the following automatic metrics for the proposed TOP task: Content Preservation Index~(CPI), Preference Preservation Index~(PPI), and Natural Realism Index~(NRI).
In this section, we will elucidate in detail the specific evaluation metrics selected for text and image modalities.

\textbf{TOP on Text.}
For TOP on text, we use \textbf{Rouge-L}~\cite{rouge}, \textbf{Bert-Score}~\cite{zhang2019bertscore} as the CPI and PPI.
For the NRI, we use the \textbf{perplexity}~(PPL) to evaluate the quality of the generated texts.

\textbf{TOP on Image.}
Following previous research~\cite{ye2023ip, ruiz2023dreambooth, deng2022stytr2}, we use \textbf{$\mathcal{L}_s$} and \textbf{$\mathcal{L}_c$} as CPI and PPI for style transfer, \textbf{SSIM}~\cite{ssim} and \textbf{CLIP-I}~\cite{ruiz2023dreambooth} as the CPI and PPI for concept transfer.
For the NRI, we use the \textbf{AES-Score}~(aesthetic quality predicted by LAION~\cite{schuhmann2022laion}) to evaluate the quality of the generated images.

\subsection{Experimental Details}
In this subsection, we introduce in detail the models of LLMs and LVMs we utilized in basic implementation.

For the TOP on text, we utilize GPT-4-32k to extract the user preference and generate customized introductions in the dataset creation.
In the training stage, we utilize the ChatGLM-v2~\cite{zeng2022glm} as the backbone model and employ p-tuning~\cite{liu-etal-2022-p} for fine-tuning.

For the TOP on image, we utilize Stable Diffusion~(SD-v1.5)\footnote{https://huggingface.co/runwayml/stable-diffusion-v1-5}, OpenCLIP ViT-H/14~\cite{clip}, and ViT-B/16~\cite{dosovitskiy2020vit}.

\subsection{Main Results}
The experimental results presented in Table~\ref{tab:result:text} are derived from the implementation approach selected for the TOP framework discussed in this paper. 
The significance of these results lies in providing reference baseline results and indicating future directions for improvement within the TOP task, rather than representing the ultimate performance.

The results of the text indicate that under the existing implementation, the performance metrics for CPI and NRI exhibit relatively good outcomes, whereas the PPI metric is not very well.
Besides, the results on images demonstrate that the TOP framework achieves good performance for style transfer, but its performance in concept transfer is relatively average.

Figure~\ref{fig:teaser:result:result} and Figure~\ref{fig:teaser:result:concept} illustrate the teasers of the TOP on image.

%% file: section/7.conclusion.tex
\section{Conclusion}
In this paper, we propose a new \task~(TOP) task.
The TOP task aims to generate more customized outputs depending on user preferences.
We provide the basic concepts and framework definitions of the TOP task and provide the basic implementation of the TOP framework based on LLMs and LVMs on text and image modalities.
We also release a related dataset Customized-TOP which includes $46,372$ texts and $39,685$ images.
To better measure the performance of the model in the TOP task, we give three types of metrics that should exist in TOP framework: CPI, PPI and NRI, and give the specifically used metrics respectively.
Finally, we give the experimental results of the basic implementation on Customized-TOP as a reference baseline.